\documentclass[fleqn,usenatbib]{mnras}

\usepackage{newtxtext,newtxmath}

\usepackage[T1]{fontenc}

\DeclareRobustCommand{\VAN}[3]{#2}
\let\VANthebibliography\thebibliography
\def\thebibliography{\DeclareRobustCommand{\VAN}[3]{##3}\VANthebibliography}

\usepackage{graphicx}	
\usepackage{amsmath}	

\usepackage{amsmath}
\usepackage{hyperref}
\usepackage{rotating}
\usepackage{longtable}
\usepackage{multirow}
\usepackage[normalem]{ulem}
\usepackage[flushleft]{threeparttable}
\usepackage{bm}
\usepackage{makecell}
\usepackage{array}

\usepackage{gensymb}
\usepackage{ragged2e}
\usepackage{graphicx}	
\usepackage{amsmath}	
\usepackage{xcolor}
\definecolor{seagreen}{rgb}{0.18, 0.55, 0.34}

\usepackage{comment}
\usepackage{soul}

\title[Detection of sGRBs from BNS mergers]{Joint gravitational wave-short GRB detection of Binary Neutron Star mergers with existing and future facilities}

\author[Bhattacharjee et al.]{
Soumyadeep Bhattacharjee,$^{1,2}$
Smaranika Banerjee,$^{3}$
Varun Bhalerao$^{4}$,
Paz Beniamini $^{5,6,7}$,\newauthor
Sukanta Bose $^{8}$,
Kenta Hotokezaka $^{9}$,
Archana Pai $^{4}$,
Muhammed Saleem $^{10}$,
and
Gaurav Waratkar $^{4}$
\\
$^{1}$Department of Physics, Indian Institute of Science, Bengaluru, 560012, India\\
$^{2}$Department of Astronomy, California Institute of Technology, 1200E. California Blvd, Pasadena, CA, 91125, USA\\
$^{3}$The Oskar Klein Centre, Department of Astronomy, Stockholm University, AlbaNova, SE-10691 Stockholm, Sweden\\
$^{4}$Department of Physics, IIT Bombay, Powai, Mumbai 400076, India\\
$^{5}$Astrophysics Research Center of the Open University (ARCO), The Open University of Israel, P.O Box 808, Ra’anana 4353701, Israel\\
$^{6}$Department of Natural Sciences, The Open University of Israel, P.O Box 808, Ra’anana 4353701, Israel \\
$^{7}$Department of Physics, The George Washington University, 725 21st Street NW, Washington, DC 20052, USA\\
$^{8}$Department of Physics and Astronomy, Washington State University, Webster Hall Pullman, WA 99163, USA\\
$^{9}$Reaseach Center for the Early Universe, the University of Tokyo, Hongo 7-3-1, Bunkyo, Tokyo 113-0033, Japan\\
$^{10}$School of Physics and Astronomy, University of Minnesota, Twin cities, Minneapolis, MN 55455\\
}
\date{Accepted XXX. Received YYY; in original form ZZZ}

\pubyear{2015}

\begin{document}
\label{firstpage}
\pagerange{\pageref{firstpage}--\pageref{lastpage}}
\maketitle

\begin{abstract}
We explore the joint detection prospects of short gamma-ray bursts (sGRBs) and their gravitational wave (GW) counterparts by the current and upcoming high-energy GRB and GW facilities from binary neutron star (BNS) mergers. We consider two GW detector networks: (1) A four-detector network comprising LIGO Hanford, Livingston, Virgo, and Kagra, (IGWN4) and (2) a future five-detector network including the same four detectors and LIGO India (IGWN5). For the sGRB detection, we consider existing satellites \emph{Fermi} and \emph{Swift} and the proposed all-sky satellite \emph{Daksha}. Most of the events for the joint detection will be off-axis, hence, we consider a broad range of sGRB jet models predicting the off-axis emission. Also, to test the effect of the assumed sGRB luminosity function, we consider two different functions for one of the emission models. We find that for the different jet models, the joint sGRB and GW detection rates for \emph{Fermi} and \emph{Swift} with IGWN4 (IGWN5) lie within 0.07-0.62$\mathrm{\ yr^{-1}}$ (0.8-4.0$\mathrm{\ yr^{-1}}$) and 0.02-0.14$\mathrm{\ yr^{-1}}$ (0.15-1.0$\mathrm{\ yr^{-1}}$), respectively, when the BNS merger rate is taken to be 320$\mathrm{\ Gpc^{-3}~yr^{-1}}$. With \emph{Daksha}, the rates increase to 0.2-1.3$\mathrm{\ yr^{-1}}$ (1.3-8.3$\mathrm{\ yr^{-1}}$), which is 2-9 times higher than the existing satellites. We show that such a mission with higher sensitivity will be ideal for detecting a higher number of fainter events observed off-axis or at a larger distance. Thus, \emph{Daksha} will boost the joint detections of sGRB and GW, especially for the off-axis events. Finally, we find that our detection rates with optimal SNRs are conservative, and noise in GW detectors can increase the rates further.
\end{abstract}




\section{Introduction}

\label{sec:intro}
Binary neutron star (BNS) mergers have long been hypothesized to be associated with short-duration gamma-ray bursts (sGRBs).
This is observationally confirmed for the first time by the joint detection of gravitational waves (GW170817, \citealt{Abbott17})
and the spatially coincident sGRB (GRB170817A, \citealt{Savchenko17, Goldstein17}) from a BNS merger.
Except for the understanding of the sources of the sGRBs, the extensive multi-wavelength follow-up observations of the afterglow significantly improves
our understanding of the sGRB jets and emission processes \citep{Troja17,Troja18,Margutti18,Lyman18,Alexander18,Mooley18a,Mooley18b,Ghirlanda19,Troja20,Beniamini2020,Beniamini2022}. 

An interesting aspect of the joint detection of GW and sGRB is that the exact time of the merger is known.
Hence, much information about the jet (such as, jet launching time) and the propagation of the jet through the ejecta can be analyzed.
Therefore, joint detection of sGRB and GW from BNS mergers is the ideal probe to understand not only the sources of the sGRBs
but also the mechanism of the jet and the emission \citep{Nakar07,Berger14,Nakar20}.
Unfortunately, after the joint detection of GRB170817A and GW170817, there has been no other confirmed joint sGRB and GW detection, which hinders advancing our understanding of the sGRB sources and the jet mechanism.

Joint GW and sGRB detections depend on the sensitivities and capabilities of both the GW detectors and the satellites.
Currently, the fourth observation (O4) run of the LIGO-Virgo-KAGRA collaboration is ongoing~(\citealp{Aasi15,Acernese15,Aso13}). In the future, LIGO detectors at Hanford and Livingston will be upgraded to A+ sensitivity \citep{Barsotti18}, significantly improving their performance.
Moreover, the planned GW detector LIGO-India~(\citealp{Iyer11,Saleem22}) will also join the detector network, further improving the prospects of GW detection.

As the sensitivity and range of GW detectors increase significantly in the 
future, it is likely that the bottleneck for joint detections will soon be the sensitivity of EM detectors like \emph{Fermi} and \emph{Swift}. Thus, more sensitive high-energy satellites are required to improve the prospects of joint detections.

One such satellite proposed to study explosive astrophysical sources like GRBs and other high-energy electromagnetic counterparts to GW sources is \emph{Daksha} (\citealp{Bhalerao22a,Bhalerao22b}).
The Daksha mission will have two high-energy space telescopes with three types of detectors, covering the broad energy range from 1~keV to $\sim1~$MeV. The high sensitivity of Daksha and its near-uniform all-sky sensitivity arise from the Cadmium Zinc Telluride (CZT) detectors covering the medium energy range from $20-200\mathrm{~keV}$, with a median 
effective area of $\sim1310\mathrm{~cm^2}$. Thus, \emph{Daksha} has an effective area significantly higher than \emph{Fermi}-GBM and will achieve \emph{Swift}-BAT like sensitivity over the entire sky.
With this high sensitivity, \emph{Daksha} will clearly be a formidable instrument in the search for and study of high-energy transients.

In this paper, we study the prospects of joint sGRB and GW detection by the existing missions \emph{Fermi} and \emph{Swift} and the proposed \emph{Daksha} mission. We consider multiple sGRB emission models to probe the various proposed models for the prompt emission. There have been discussions about sGRB-like emissions from neutron star--black hole (NSBH) mergers as well, however the theoretical details are still murky, and there is no observational confirmation. Hence, we do not consider NSBH mergers in this study. The plan of the paper is as follows: In \S\ref{sec:sources_detectors}, we provide details of  the high-energy satellites and
the synthetic BNS population employed here. In \S\ref{sec:models},
we describe the various prompt emission models considered in this work, followed by the method used to calculate the flux for the satellites in \S\ref{sec:sims}. In \S\ref{sec:sgrb_det},
we compare the properties of different models and calculate the sGRB detection rates for the three satellites. \S\ref{sec:emgw_detection}
presents the rates and prospects of joint sGRB and GW detection. Finally, in \S\ref{sec:conclusion}, we discuss our main results, compare them with previous works, and present our conclusion.

\section{Methods}

To study the prospects of the joint sGRB and GW detection from BNS mergers,
we inject the sources in comoving volume. We then calculate the detection probabilities of the GW and GRB detectors, both separately and jointly. To probe the sensitivity of joint detection rates on the specifics of the GRB jet, we consider a broad range of sGRB jet models.
Note that most of the candidate sources for the joint detection will be observed off-axis. Hence, in this work, we limit ourselves only to a set of models that make quantitative predictions for off-axis emission.

\subsection{The Sources and the Detectors}\label{sec:sources_detectors}

\begin{figure}
    \centering
    \includegraphics[width=\linewidth]{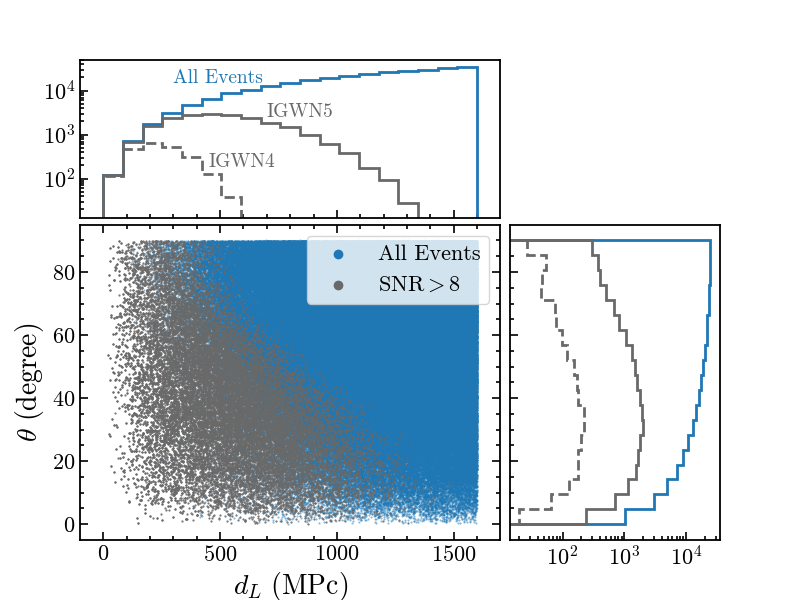}
    \caption{The distribution of the injected BNS merger events over viewing angle, $\theta$, and luminosity distance $d_L$ (blue points). The histograms show the corresponding distributions (blue histogram). In the scatter plot, we also show the distribution of the events which are detected over threshold SNR of 8 either by IGWN4 or IGWN5 network (grey points). In the histograms, we show the distribution of these events individually for IGWN4 (dashed grey histogram) and IGWN5 (solid grey histogram).}
    \label{fig:source_props}
\end{figure}

\begin{table*}

\caption{The energy bands, flux thresholds, and effective detection fraction of the high-energy missions. $f_{\rm sky}$ denotes the fraction of sky which is not earth-occulted for the satellite, and $f_{\rm not\_SAA}$ denotes the fraction of the satellite's orbit inside SAA. $f_{\rm df}$, thus, denotes the fraction of merger events that are not missed by the satellite for either being earth-occulted or for SAA-outage. }
\label{tab:det_frac}
\begin{threeparttable}
\begin{tabular}{|c|c|c|c|c|c|}
\hline
\multirow{2}{*}{Mission} & Energy band & Detection Threshold & Sky-coverage & SAA-outage fraction & Total detection fraction \\
& $\left(\mathrm{keV}\right)$ & $\left(10^{-8}\mathrm{\ erg \ s^{-1}}\right)$ & $\left(f_{\mathrm{sky}}\right)$
& $\left(1-f_{\mathrm{not\_SAA}}\right)$ & $\left(f_{\mathrm{df}} = f_{\mathrm{sky}}\times f_{\mathrm{not\_SAA}}\right)$ \\

\hline
Daksha (two)$^\text{(a)}$ & 20-200 & 4 & 1            & 0.134               & 0.866                    \\
Fermi-GBM$^\text{(b)}$    & 50-300 & 3 & 0.7          & 0.15                & 0.595                    \\
Swift-BAT$^\text{(c)}$    & 15-150 & 20 & 0.11         & 0.10                & 0.100   \\    
\hline
\end{tabular}
\begin{tablenotes}
      \item \textbf{References: }(a) \citet{Bhalerao22a,Bhalerao22b} (b) \citet{vKienlin20,Poolakkil21}, \url{https://fermi.gsfc.nasa.gov/science/instruments/table1-2.html} (c) \citet{Barthelmy05}, \citet{Lien14}, \url{https://swift.gsfc.nasa.gov/about_swift/bat_desc.html}, \url{https://swift.gsfc.nasa.gov/analysis/bat_digest.html}
     \end{tablenotes}
  \end{threeparttable}
\end{table*}

We inject BNS merger sources up to a luminosity distance of $d_L=1.6\mathrm{\ GPc}$
(comoving distance of $\approx$1.2~Gpc and $z\sim0.3$\footnote{Flat cosmology with $H_0=67.7\,\mathrm{km\,s^{-1}\,Mpc^{-1}}$ and $\Omega_{\mathrm{M}}=0.31$, \citet{Planck13}}). We distributed the injected sources uniformly in the source-frame comoving volume, which translates to a redshift probability distribution of
\begin{equation}\label{eq:source_frame}
    \frac{dp(z)}{dz} \propto \frac{1}{1+z}\frac{dV_c}{dz}.
\end{equation}
The inclinations of the sources are isotropically distributed. The masses of the neutron stars in the sources are drawn from a normal distribution with a mean of $1.33\ \mathrm{M}_\odot$ and a standard deviation of $0.09\ \mathrm{M}_\odot$ \citep{ozel16}.
The spins of the neutron stars are drawn uniformly between $0<\chi<0.05$, with the upper limit set by the maximum known NS spin \citep{Zhu18}.
The GW waveforms are generated using IMRPhenomPv2 \citep{Schmidt12,Hannam14,Khan16}.

Note that the total number of sources considered are $N_{\rm inj} = 0.3$ million, which is over-sampled by a factor of $\sim 130$ from the rate of the BNS mergers within the same comoving volume
(the median rate $\sim2316$, \citealp{Abbott21b,Abbott21c}). Such over-sampling is done to avoid any effect of small number statistics on our results.
Figure \ref{fig:source_props} shows the distribution of the sources with the observation angle ($\theta$) and luminosity distance ($d_L$).

To determine the joint detection probability of GW and sGRB,
we consider the detection prospects of GW sources by two different GW detector networks,
comprising of (1) International Gravitational Wave Network-4 or IGWN4 with four detectors:
Advanced LIGO Hanford and Livingston \citep{Aasi15}, Advanced Virgo \citep{Acernese15}, and design sensitivity Kagra \citep{Aso13} and 
(2) International Gravitational Wave Network-5 or IGWN5 with five detectors:
A+ sensitivities of LIGO Hanford and Livingston \citep{Barsotti18}, A+ sensitivity LIGO-India \citep{Iyer11},
Advanced Virgo and upgraded Kagra. We use the publicly available\footnote{\url{https://git.ligo.org/sensitivity-curves/observing-scenario-paper-2019-2020-update/-/tree/master}} power spectral densities (PSDs): \texttt{aligo\_O4high.text}, \texttt{avirgo\_O4high\_NEW.text}, and \texttt{kagra\_25MPc.text} for IGWN4; and \texttt{AplusDesign.text} (also for LIGO-India), \texttt{avirgo\_O5high\_NEW.text}, and \texttt{kagra\_80MPc.text} for IGWN5. For both networks, we consider that individual detectors have an uncorrelated downtime of 30\% (following the treatment by \citealp{Petrov21}).
Our study with IGWN4 provides the prospects of joint detection in the current and near-future GW detector setup, whereas
that with IGWN5 aims to estimate the future prospects when all five detectors come online.

We calculate the optimal SNR of the sources by simulating waveforms without noise and using PSD models (see \S\ref{subsec:comparison} for a discussion on the possible effect of noise). As the criteria for the detection of GW, we consider the events above a particular threshold value of the network signal-to-noise 
($\mathrm{SNR_{net}} = \sqrt{\sum_{\rm det} \mathrm{SNR}_{\rm det}^2}$, where $\mathrm{SNR}_{\rm det}$ is the signal-to-noise ratio (SNR) of a single detector)
\citep{Pai:2000zt}. 
In this study, we consider two values of $\mathrm{SNR_{net}}$ as $\sim 8$ and $6.5$ (following \citealp{Petrov21}), and we define these $\mathrm{SNR_{net}}$ criteria as threshold and sub-threshold detections, respectively. 

To determine the prospects of sGRB detection, we consider two already existing high-energy GRB detectors:
\emph{Fermi}-GBM ($50-300\mathrm{\ keV}$) and \emph{Swift}-BAT ($15-150\mathrm{\ keV}$), and the Medium-Energy (ME) detectors of the proposed mission \emph{Daksha} ($20-200\mathrm{\ keV}$).
Note that these satellites are low earth-orbit satellites. These satellites suffer from limited sky coverage since the Earth blocks about 30$\%$ of the sky for satellites in low earth orbit. For instance, \emph{Fermi}-GBM can cover roughly about a fraction of 0.7 of the sky. In the case of \emph{Daksha}, the two satellites on opposite sides of the earth overcome this limitation, making the fraction $1$. In addition, all satellites in low earth-orbit are inactive during passage through the South Atlantic Anomaly (SAA).

We account for the effects of limited sky coverage and SAA by assigning an effective detection fraction ($f_{\mathrm{df}}$)
composed of the SAA out-time and sky-coverage fraction. To determine whether a sGRB is detectable by a satellite,
we consider the thresholds in terms of the average flux received from the source by each of the detectors.
The flux thresholds and the $f_{\mathrm{df}}$ values considered for the three satellites are listed in Table~\ref{tab:det_frac}.

\subsection{The emission models} \label{sec:models}
The sGRB emission is produced from a relativistic jet launched after the central remnant of the BNS merger collapses into a BH or hypermassive NS.
If the emission arises only from the relativistic jet inside the jet opening angle ($\theta_0$), the emission declines very rapidly beyond the jet axis. However, the jet can have angular structure either from the intrinsic structure of the jet or the interaction of the jet with the surrounding ejecta,
forming a cocoon surrounding the jet \citep{Nakar20}.
The angular structure of the jet depends on several factors, such as the jet opening angle, the jet launching time after the merger,
the initial energy of the jet, which determines whether the jet can successfully break out of the ejecta.
The angular structure of the jet makes the observations beyond the jet core possible.

In this work, we adopt only jets with angular structures beyond the jet cores since we consider the sGRB associated with the GW, which, in most cases,
will be observed off-axis. For example, we consider the jets with a Gaussian or power law distribution of the total energy ($E$ or luminosity, $L$), the energy at the spectral peak ($E_{\rm peak}$), and/or the Lorentz factor ($\Gamma$) with the angle from the jet axis (hereafter structured jets) and the jets with cocoon structure beyond the jet core (hereafter jet-cocoon). Note that all these models considered here fall under the broad category of structured jets \citep{Nakar20}.
However, in this work, we define the models separately for convenience.

\subsubsection{Structured Jets}\label{subsec:mod_plj_gaussian}

We adopt two structured jet models. In the first model, the beaming-corrected isotropic energy in the bolometric band, $E_{\mathrm{iso}}^{\mathrm{bol}}$,  varies as follows: $E_{\mathrm{iso}}^{\mathrm{bol}}$ is assumed to be constant within the jet core ($E_{\mathrm{iso}}^{\mathrm{bol}}(\theta) = E_0$ at $\theta < \theta_0$), beyond which it varies as a power law as $E_{\mathrm{iso}}^{\mathrm{bol}}(\theta) = E_0(\theta/ \theta_{0})^{-\delta}$. Furthermore, we also assume that the Lorentz factor distribution has a similar angular profile with $\Gamma_0$ at the core.
We call this model the "Power-Law Jet" (henceforth PLJ, note that we name the models after the type of angular dependence of the energy for ease of reference). We follow the treatment of \citet{Beniamini19a,Beniamini19b} for the model parameters: $\theta_{0} = 0.1\mathrm{\ rad}\approx5.7\degree$ and $\delta = 4.5$. The $\log_{10}(E_{\mathrm{peak,c}})$ distribution along the core is drawn from a normal distribution with a median value of 2.7 and a standard deviation of 0.19 \citep{Nava2011}. We then vary the value of $E_{\mathrm{peak}}$ with $\theta$ such that $E_{\mathrm{peak}}(\theta)=E_{\mathrm{peak,c}} (\Gamma(\theta)/\Gamma_0)$, i.e. considering a situation in which the value of the peak energy remains constant in the co-moving frame. We note that when the observed emission is dominated by $\theta\ll \theta_{\rm obs}$, then due to relativistic Doppler beaming, we have $E_{\rm peak,obs}=E_{\rm peak}(\theta) \left[(\theta_{\rm obs}-\theta)\Gamma(\theta)\right]^{-2}$. The event duration was drawn randomly from a distribution of cosmological sGRB \citep{BDPG2020}.
All the above jet parameters are defined in the observer frame.

We consider another structured jet model with a Gaussian variation of the radiative energy with angle: $E_{\gamma}(\theta) = E_0e^{-\theta^2/2\theta_0^2}$, following the works of \citet{Ioka19}. Unlike the previous model, the parameters here are defined in the comoving frame. It further assumes a power variation law of the Lorentz factor: $\Gamma(\theta) = \Gamma_{\rm max}/(1 + (\theta/\theta_{0})^{\lambda})$. Following the treatment of \citet{Ioka19}, we use $\theta_0=0.059\mathrm{\ rad}\approx3.4\degree$. Appropriate relativistic beaming corrections are then applied to obtain the beaming-corrected isotropic emission profile in the observer frame. For the details of the calculation, refer to \S3 of their paper. Based on our convention, we call this model the "Gaussian."

\subsubsection{Jet-Cocoon model}\label{subsec:mod_cocoon}
We adopt the Jet-Cocoon estimates of \citep{Beniamini19a}. The model consists of a top hat jet (with constant bolometric luminosity $L_{0}$ within the jet core angle $\theta_{0}$) surrounded by a quassi-spherical cocoon emission with bolometric isotropic equivalent luminosity $L_{\gamma,\rm co}$. Hence the cocoon model will follow $L_{\mathrm{iso}}^{\mathrm{bol}}(\theta) = L_{0} + L_{\gamma,\rm co}$ for $\theta_{\rm obs}<\theta_{0}$; and $L_{\mathrm{iso}}(\theta) = L_{\gamma,\rm co}$ for $\theta_{\rm obs}>\theta_{0}$, where $\theta_0=0.1\mathrm{\ rad}\approx5.7\degree$. The value of the jet core's $E_{\mathrm{peak}}$ is taken to be the same as described above for the PLJ model. The value corresponding to the cocoon is taken as $100$\,keV. The jet engine durations were taken as mentioned before for the structured jets.
For more details, see \citet{Beniamini19a}. We call this model the "Jet-Cocoon" model. 
The interaction between the jet and the ejecta may result in some low-energy jets failing to break out of the ejecta. We refer to such events as ``failed jet" events. This results in very low luminosity values, rendering them mostly undetectable. Under fiducial simulation conditions, about $1/3^{\rm rd}$ of the jets fail in the estimates of \citet{Beniamini19a}. For the details of the jet-ejecta interaction parameters, refer to the paper. We denote the fraction of events having successful jets as $f_{\rm jet}$, which, in this work, is $2/3$.

  
\subsection{Flux Calculation} \label{sec:sims}

For each of the injected events, we calculate the flux received by the different satellites (e.g., the existing satellites \emph{Swift} and \emph{Fermi} and the upcoming satellite \emph{Daksha})
within their respective energy bands. The flux received is calculated as $F = L_{\mathrm{iso}}^{\mathrm{sat}}/4\pi d_L^2$,
where $L_{\mathrm{iso}}^{\mathrm{sat}}$ is the isotropic luminosity within the energy band of the satellite, and $d_L$ is the luminosity distance to the source.
For all the models except Gaussian, we calculate the energy in the respective energy bands of the satellites ($L_{\mathrm{iso}}^{\mathrm{sat}}$) as:
\begin{equation}
    L_{\mathrm{iso}}^{\mathrm{sat}}(\theta) = \frac{\int_{E_1(1+z)}^{E_2(1+z)}EN(E)dE}{\int_0^\infty EN(E)dE}L_{\mathrm{iso}}^{\mathrm{bol}}(\theta),
\end{equation}
where $E_1$ and $E_2$ are the energy band limits and $z$ is the source redshift. To calculate $ N(E)$, we assume a comptonized spectral model with a cutoff power law:
\begin{equation}\label{eq:cutoff_pl_spectrum}
N(E, \theta) \propto E^\alpha \exp\left(-\frac{E}{E_c(\theta)}\right) ,
\end{equation}
where $E_c$ is the cut-off energy defined as: $E_c = E_{\mathrm{peak}}/(2 + \alpha)$ and $\alpha$ is the power-law index.
We use $\alpha=-0.6$, following \citet{Abbott17} for the analysis of GRB170817A, and the value of $E_{\mathrm{peak}}$ at different angles are given by the models (see \ref{sec:models}).

The Gaussian model, instead, uses a Band-function \citep{Band93} like spectral energy distribution in the comoving frame:
\begin{equation}
    f(E,\theta) \propto \frac{1}{E_0(\theta)}\left(\frac{E}{E_0}\right)^{1+\alpha}\left[1+\left(\frac{E}{E_0(\theta)}\right)^2\right]^{\frac{\beta-\alpha}{2}}
\end{equation}
where $E_0(\theta) = 0.15\mathrm{\ keV}(1+(\theta/\theta_0)^{0.75})$ with $\theta_0 = 0.059\mathrm{\ rad}$. The values of $\alpha$ and $\beta$ are taken to be $1$ and $2.5$, respectively \citep{Kaneko06}. The spectral distribution is used in conjunction with the angular distribution of energy and Lorentz factor (see \S\ref{subsec:mod_plj_gaussian}) to determine the beaming corrected emission energy within a desired energy range, which is given by Equation 14 in \citet{Ioka19}. We use this equation to calculate the energy received by the different satellites within their respective energy bands.

The $L_{\rm iso}$ at the jet axis ($\theta=0$, $L_0$) has been drawn from the intrinsic on-axis energy (or luminosity)
distribution of sGRBs given by \citet[][hereafter WP15]{WP15}.
The distribution (called the luminosity function) takes the form of a broken power law:
\begin{equation}
    \phi\left(L_{0}\right) = \frac{dN_{\rm GRB}}{d\log(L_{0})} = 
    \begin{cases}
      \left(\frac{L_{0}}{L_*}\right)^{-\alpha_L} & L_{0}<L_*\\
      \left(\frac{L_{0}}{L_*}\right)^{-\beta_L} & L_{0}>L_*,
    \end{cases} 
\end{equation}
where, $\alpha_L=1$, $\beta_L=2$ and $L_*=2\times10^{52}\mathrm{\ erg~s^{-1}}$.

Note that the uncertainty in the assumed luminosity function may affect the predicted rates.
To study the sensitivity of our results to the choice of the luminosity function, we also use a different luminosity function taking the Gaussian jet as the representative model.
For this purpose, we use the isotropic energy distribution of the observed \emph{Swift} GRBs as reported in \citet[][hereafter F15]{Fong15}. 
The distribution takes the form of a Gaussian in the log space with center at $\log(E_{0}) = 51.31\mathrm{\ erg~s^{-1}}$ and a standard deviation of $0.98$. To calculate the corresponding luminosity, we assume a jet activity time of $0.3$ seconds to calculate the luminosity for all the events, which corresponds to typical sGRB duration \citep{Kouveliotou93}. Hereafter, we call this model as the Gaussian-F model.
The distribution is biased towards higher energy values since this is designed from the isotropic energy distribution of the observed sGRBs, unlike the WP15 function.

After calculating the flux, we calculate the detection rates of prompt emission by various satellites and also the joint detection rates based on various criteria (see following sections). In the jet-cocoon model we have to explicitly account for unsuccessful jets as the cocoon emission might still be detectable in those cases. Therefore we multiply the calculated detection rates with the structured jet models by $f_{\rm jet}=2/3$, i.e. the number of successful jet break-outs is assumed to be the same in the different models. This provides a fair comparison between Jet-Cocoon and Structured Jet models.

\section{Detection of the prompt emission}\label{sec:sgrb_det}
\begin{table*}
\caption{The sGRB and sGRB and GW joint detection rates for all GW network, mission, and emission model combinations. The above-threshold GW detection rates for BNS events are 15~yr$^{-1}$ and 148~yr$^{-1}$ for IGWN4 and IGWN5 respectively, with corresponding sub-threshold rates of 26~yr$^{-1}$ and 264~yr$^{-1}$. The three ``sGRB detection rate'' columns give the EM rates for the three missions, independent of GW detection. The last six columns give joint detection rates. For instance, the last cell of the table states that for the Gaussian-F model with the F15 luminosity function, \emph{Daksha} + IGWN5 will detect 8.27 above-threshold events and another 13.22 sub-threshold events per year, and 94\% of the above-threshold events will be off-axis events.}
\label{tab:rate_table}
\centering
\resizebox{\linewidth}{!}{%
\begin{tabular}{|c|c|c|c|c|c|c|c|c|c|c|} 
\hline
Models     
& \begin{tabular}[c]{c}Luminosity\\function\end{tabular} 
& \multicolumn{3}{c|}{\begin{tabular}[c]{@{}c@{}}sGRB detection rate\\$(\mathrm{yr^{-1}})$\end{tabular}} 
& \multicolumn{3}{c|}{\begin{tabular}[c]{@{}c@{}}IGWN4 joint detection rates (yr$^{-1}$)\\Threshold, Sub-Threshold\\$f_{\mathrm{off}}$ (\%)\end{tabular}} 
& \multicolumn{3}{c|}{\begin{tabular}[c]{@{}c@{}}IGWN5 joint detection rates (yr$^{-1}$)\\Threshold, Sub-Threshold\\$f_{\mathrm{off}}$ (\%)\end{tabular}}  \\ 
\hline
           &                                                              & Fermi & Swift & Daksha                                                                                 & Fermi                                                    & Swift                                                    & Daksha                                                                   & Fermi                                                    & Swift                                                    & Daksha                                                                       \\ 
\hline

Jet-Cocoon & WP15                                                         & 3.55  & 0.64  & 5.52                                                                                   & \begin{tabular}[c]{@{}c@{}}0.07, 0.13\\41\%\end{tabular}  & \begin{tabular}[c]{@{}c@{}}0.02, 0.03\\59\%\end{tabular} & \begin{tabular}[c]{@{}c@{}}0.17, 0.30\\53\%\end{tabular}                 & \begin{tabular}[c]{@{}c@{}}0.8, 1.48\\13\%\end{tabular}  & \begin{tabular}[c]{@{}c@{}}0.15, 0.26\\19\%\end{tabular} & \begin{tabular}[c]{@{}c@{}}1.25, 2.24\\16\%\end{tabular}                     \\

PLJ        & WP15                                                         & 5.64  & 2.54  & 19.54                                                                                  & \begin{tabular}{c}0.38, 0.62\\81\%\end{tabular} & \begin{tabular}[c]{c}0.12, 0.21\\91\%\end{tabular} & \begin{tabular}[c]{@{}c@{}}0.95, 1.57\\88\%\end{tabular}                 & \begin{tabular}[c]{@{}c@{}}2.04, 2.81\\66\%\end{tabular} & \begin{tabular}[c]{@{}c@{}}0.83, 1.25\\81\%\end{tabular} & \begin{tabular}[c]{@{}c@{}}6.30, 9.71\\79\%\end{tabular}                     \\

Gaussian   & WP15                                                         & 1.88  & 0.94  & 7.77                                                                                  & \begin{tabular}[c]{@{}c@{}}0.21 0.30\\83\%\end{tabular} & \begin{tabular}[c]{@{}c@{}}0.05, 0.09\\89\%\end{tabular} & \begin{tabular}[c]{@{}c@{}}0.45, 0.76\\89\%\end{tabular}                 & \begin{tabular}[c]{@{}c@{}}0.89, 1.19\\77\%\end{tabular} & \begin{tabular}[c]{@{}c@{}}0.35, 0.50\\84\%\end{tabular} & \begin{tabular}[c]{@{}c@{}}2.89, 4.22\\83\%\end{tabular}                     \\ 
\hline
Gaussian-F & F15                                                          & 13.58 & 3.48  & 29.12                                                                                 & \begin{tabular}[c]{@{}c@{}}0.62, 1.03\\94\%\end{tabular} & \begin{tabular}[c]{@{}c@{}}0.14, 0.24\\96\%\end{tabular} & \begin{tabular}[c]{@{}c@{}}1.26, 2.05\\96\%\end{tabular}                 & \begin{tabular}[c]{@{}c@{}}4.00, 6.32\\93\%\end{tabular} & \begin{tabular}[c]{@{}c@{}}1.00, 1.55\\94\%\end{tabular} & \begin{tabular}[c]{@{}c@{}}8.27, 13.22\\94\%\end{tabular}                   \\
\hline
\end{tabular}
}

\end{table*}

\subsection{The model dependence of sGRB luminosity}

\begin{figure*}
    \centering
    \includegraphics[width=\linewidth]{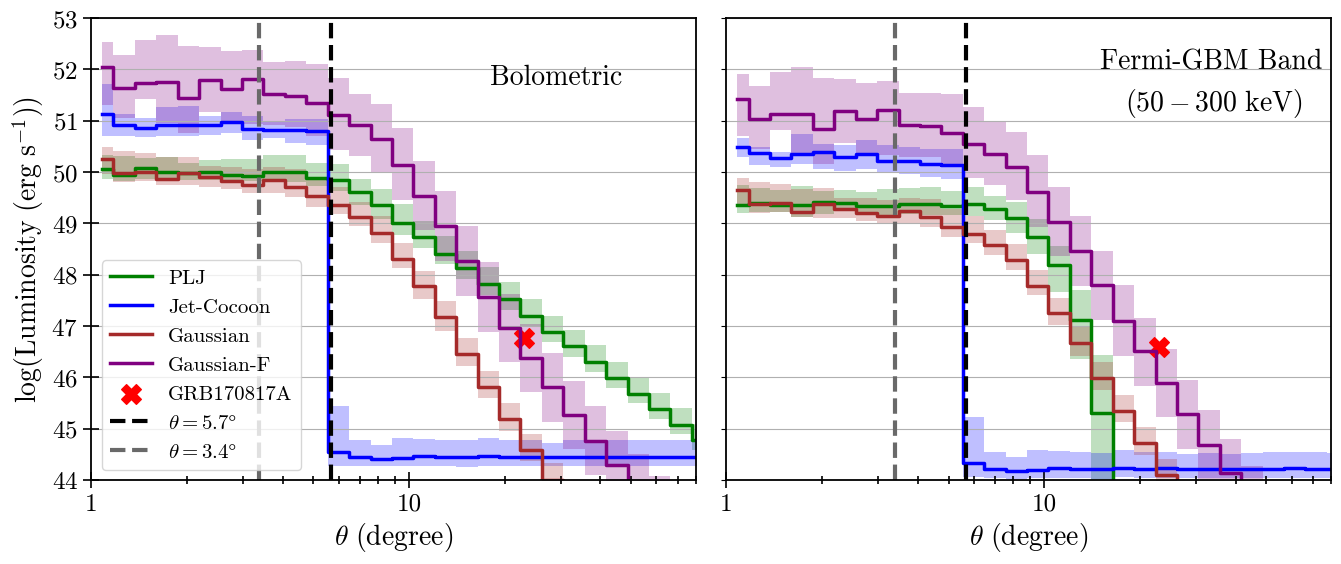}
    \caption{The luminosity ranges for all the prompt emission models considered in this work within the bolometric (\textbf{Left}) and \emph{Fermi}-GBM (\textbf{Right}) energy bands. The line represents the median luminosity, whereas the band represents the interquartile spread. GRB170817A is marked with a red cross. The jet opening angles of the Gaussian model ($0.059 \mathrm{\ rad}\approx3.4\degree$) and the PLJ and Jet Cocoon models ($0.1 \mathrm{\ rad}\approx5.7\degree$) are also marked (vertical lines grey and black lines, respectively).}
    \label{fig:flux_theta_models_fermi}
\end{figure*}

To calculate the EM detection rates by various satellites, we need to calculate the source fluxes in the appropriate energy bands. We begin by calculating the bolometric luminosities of the injected events using the models described above (Figure \ref{fig:flux_theta_models_fermi}, left panel). Then we use the source spectrum and source redshift to calculate the flux for each satellite. As an illustration, the right panel of Figure \ref{fig:flux_theta_models_fermi} shows fluxes of all simulated sources in the \emph{Fermi} GBM (50-300~kev) band.

The luminosities of the structured jets (PLJ and Gaussian) are high at smaller angles, reaching the value of $\sim10^{50}~\mathrm{erg}$, which declines with the observer angles. The luminosities inside the jet opening angle are similar for both the models, owing to their intrinsically similar energy structure inside the jet-core. However, beyond the jet-core, the PLJ is brighter than Gaussian Jet as a consequence of the assumed angular energy distributions for the two models. 

For the Jet-Cocoon model, the total gamma-ray energy of the jet and the cocoon differ by $\sim 4$ orders of magnitudes (Figure \ref{fig:flux_theta_models_fermi}). This reflects the energy difference between the jet and the cocoon. Moreover, the cocoon shows a nearly constant luminosity over viewing angles since it has nearly isotropic energy distribution. Note that although we consider both the successful and failed jets for this model, the results shown in Figure~\ref{fig:flux_theta_models_fermi} include only the successful jets. This is due to the fact that if the jet is failed and the cocoon is the source of the emission, the luminosities across all angles are extremely low ($\lessapprox10^{45}~\mathrm{erg~s^{-1}}$), making them undetectable by the different facilities (e.g., Fermi) for typical source distances.

The median on-axis luminosity of the jet-cocoon model is higher than that of the structured jet models (PLJ and Gaussian, left panel of Figure~\ref{fig:flux_theta_models_fermi}). 
This is a selection effect. As discussed before, weak jets fail to break out of the ejecta, creating very faint emissions. We discard these as mostly non-detectable. Hence, the models shown in this plot are only ones with successful jets, which in turn means that their on-axis luminosity is higher.
For the off-axis emission, PLJ shows the weakest angular dependence among the different models (left panel of Figure~\ref{fig:flux_theta_models_fermi}). 

The Gaussian-F model has higher bolometric luminosity than the Gaussian model at smaller viewing angles since the luminosity function is brighter (left panel of Figure~\ref{fig:flux_theta_models_fermi}). Additionally, the events span a larger range of luminosity since the luminosity function is wider.  Note that the Gaussian-F model is consistent with the GRB170817A luminosity since the Gaussian model \citep{Ioka19} is designed to explain this particular sGRB, and the luminosity function is drawn from the observation of the sGRBs.

The luminosities in \emph{Fermi} energy band (right panel in Figure \ref{fig:flux_theta_models_fermi}) show similar behavior as the bolometric luminosities for most models, except for the fact that the luminosities are scaled down by a factor of $\sim2-5$. 
This is expected since we consider the luminosity in a smaller wavelength range.
However, for the PLJ model, the luminosities at larger angles are significantly different from the bolometric luminosities, showing a sharp decline with angle. This is because of the assumed $E_{\rm peak}(\theta)$ distribution for this model, which drops sharply with the viewing angle.


\subsection{EM counterpart detection rates}\label{subsec:comparing_satellites}

\begin{figure*}
    \centering
    \includegraphics[width=0.47\textwidth]{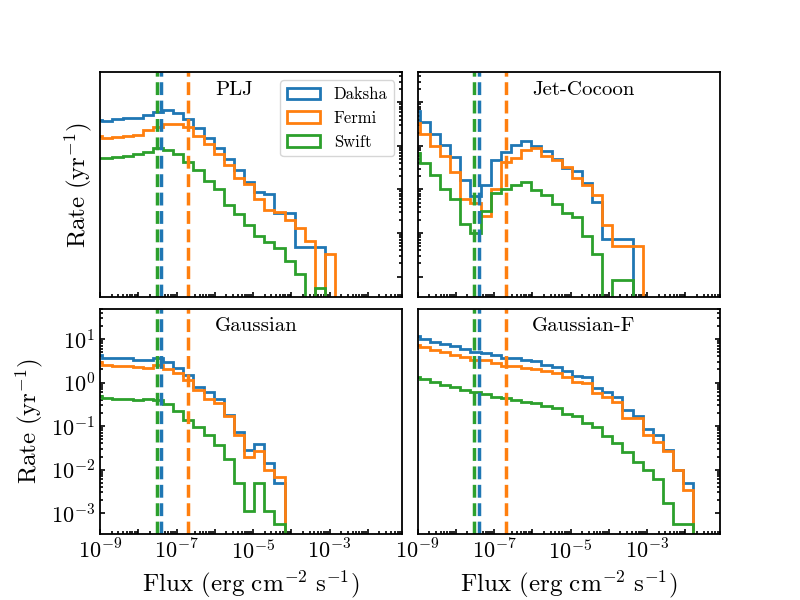}
    \hfill
    \includegraphics[width=0.47\textwidth]{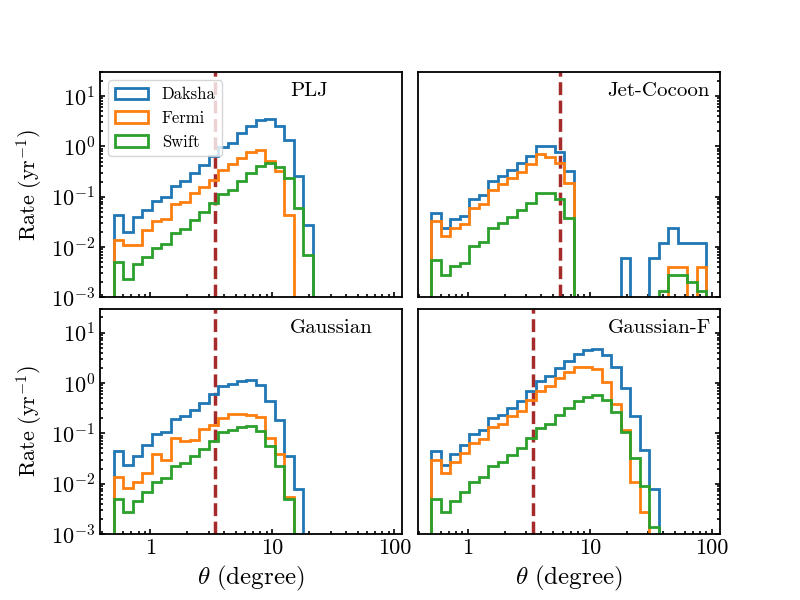}

    \caption{\textbf{Left: }The histogram of event rate per flux bin scaled by $f_{\rm df}$. In other words, the figure shows the distribution of all the events which are not missed by the satellites due to limited sky coverage or SAA outage. Vertical lines mark the detection thresholds of the satellites. \textbf{Right: }The angular distribution of the rate of sGRB detection. Vertical lines mark the jet opening angles ($\theta_0$). For all the models, \emph{Daksha} shows the highest detection rates.}
    \label{fig:flux_histogram_all_models_satellites_scaled}
\end{figure*}

In this section, we compare the sGRB detection rates for different satellites (existing \emph{Fermi} and \emph{Swift}, and upcoming \emph{Daksha}). We consider an event to be detected if the flux from the sGRB meets the detection threshold of the satellite concerned (see Table \ref{tab:det_frac} for the threshold flux values). We calculate the rates considering the median BNS merger event rate of $R_{\mathrm{BNS}}=320\mathrm{\ yr^{-1}~GPc^{-3}}$ \citep{Abbott21}. We perform a linear scaling from the number of injected events to the number of expected events within the redshifted volume (derived from Equation \ref{eq:source_frame}, see \citealt{Chen21} for a detailed discussion on volumes in this context) of the injections. To account for the limited sky coverage of the satellites and SAA outages, we perform a second scaling with a net detection fraction ($f_{\mathrm{df}}$) for a given satellite (see Table \ref{tab:det_frac}). Thus, the final relation for rate calculation becomes: 

\begin{equation}
    {\rm Rate} = f_{\mathrm{df}}R_{\rm BNS}\frac{N_{\rm det}}{N_{\rm inj}}\int_0^{z_{\rm lim}}\frac{1}{1+z}\frac{dV_c}{dz}dz,
\end{equation}

where $N_{\rm det}$ is the number of detections among the injected events, $V_c$ is the comoving volume, and $z_{\rm lim}$ is the redshift value corresponding to 1.6~Gpc. 

The detection rates of the sGRB counterparts (for BNS merger events within $d_L=1.6\mathrm{\ Gpc}$, the limit of this work) for \emph{Fermi}, \emph{Swift} and \emph{Daksha} lie within the ranges of $3.5-13.6,\ 0.6-3.5\text{, and }5.5-29.1\mathrm{\ yr^{-1}}$, respectively, for the different models. In all the cases, the rates for \emph{Daksha} are factors of $\approx2-4$ and $\approx9$ higher than \emph{Fermi} and \emph{Swift}, respectively.

The detection rates drop sharply at off-axis viewing angles (right panel of Figure~\ref{fig:dist_cum_rate_both}). For the structured jet models, the cutoff occurs in the viewing angle range of $\sim20-25\degree$, making larger viewing angle detections rare. Similar trends have also been reported in several other studies \citep{Howell19,Saleem20,Sreelekshmi}. For the Jet-Cocoon model, the cutoff occurs at $\theta\sim\theta_0$. The cocoon component, however, allows fairly large off-axis sGRB detections. Such detections, however, are only possible for nearby events ($d_L\lesssim100\mathrm{\ MPc}$). 

Figure~\ref{fig:flux_histogram_all_models_satellites_scaled} summarizes the ability of the three satellites to detect the sGRB counterparts of the injected BNS events. \emph{Swift} has the lowest detection rate among the three satellites due to its low sky coverage ($\sim10\%$). Between \emph{Fermi} and \emph{Daksha}, although the total detection rates are not much different, nevertheless Daksha has better capability for detection of the fainter (off-axis or distant) events due to its higher sensitivity and the full sky-coverage. Thus, Daksha will be able to increase the detection rates of the sGRBs, specially at the fainter ends.

We note that our analysis is limited by the luminosity distance cutoff of $1.6\mathrm{\ Gpc}$ ($z\sim0.3$). Hence, the events beyond this distance are missed by our analysis. However, in reality, many bright events with $z>0.3$ will be detected by the satellites. For instance, several sGRBs at much higher redshifts are regularly observed by \emph{Swift} and \emph{Fermi} \citep{Fong15,Poolakkil21}. Additionally, we do not consider potential sGRBs from NSBH mergers. These make the sGRB rates calculated in this work lower than the true sGRB detection rates by the satellites. This, however, is not a concern for the joint GW and sGRB detection rates, discussed in the next section, as GW detection is limited to much smaller distances with the current as well as near future detector networks.

\section{Joint GW and sGRB detection}\label{sec:emgw_detection}

\begin{figure}
    \centering
    \includegraphics[width=\linewidth]{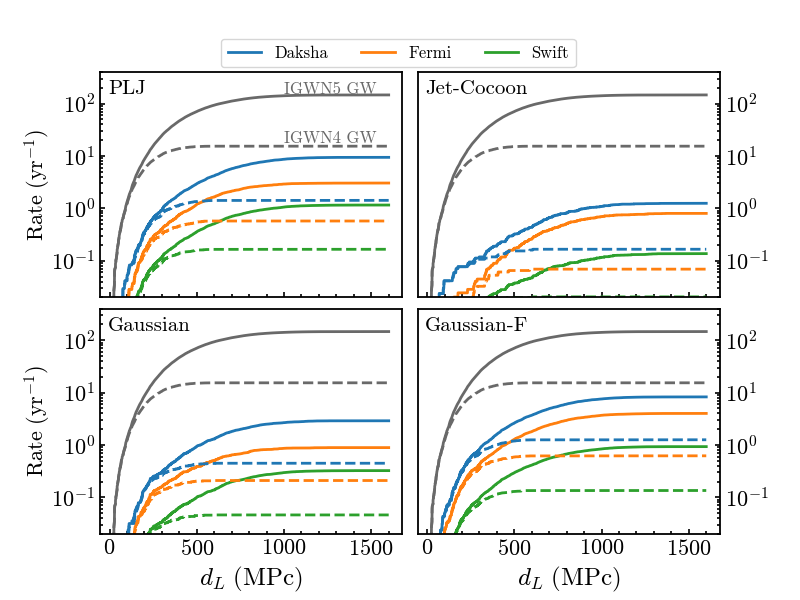}
    \caption{Cumulative joint detection rate over distance for IGWN4 (solid lines) and IGWN5 (dashed lines) networks. The grey lines show the GW detection rate, and the colored lines show joint sGRB and GW detection rates by the satellites (blue: \emph{Daksha}, green: \emph{Swift} and orange: \emph{Fermi}). The rates show a significant increase from IGWN4 to IGWN5. \emph{Daksha} performs $2-9$ times better than the existing missions.}
    \label{fig:dist_cum_rate_both}
\end{figure}

\begin{figure*}
    \centering
    \includegraphics[width=\linewidth]{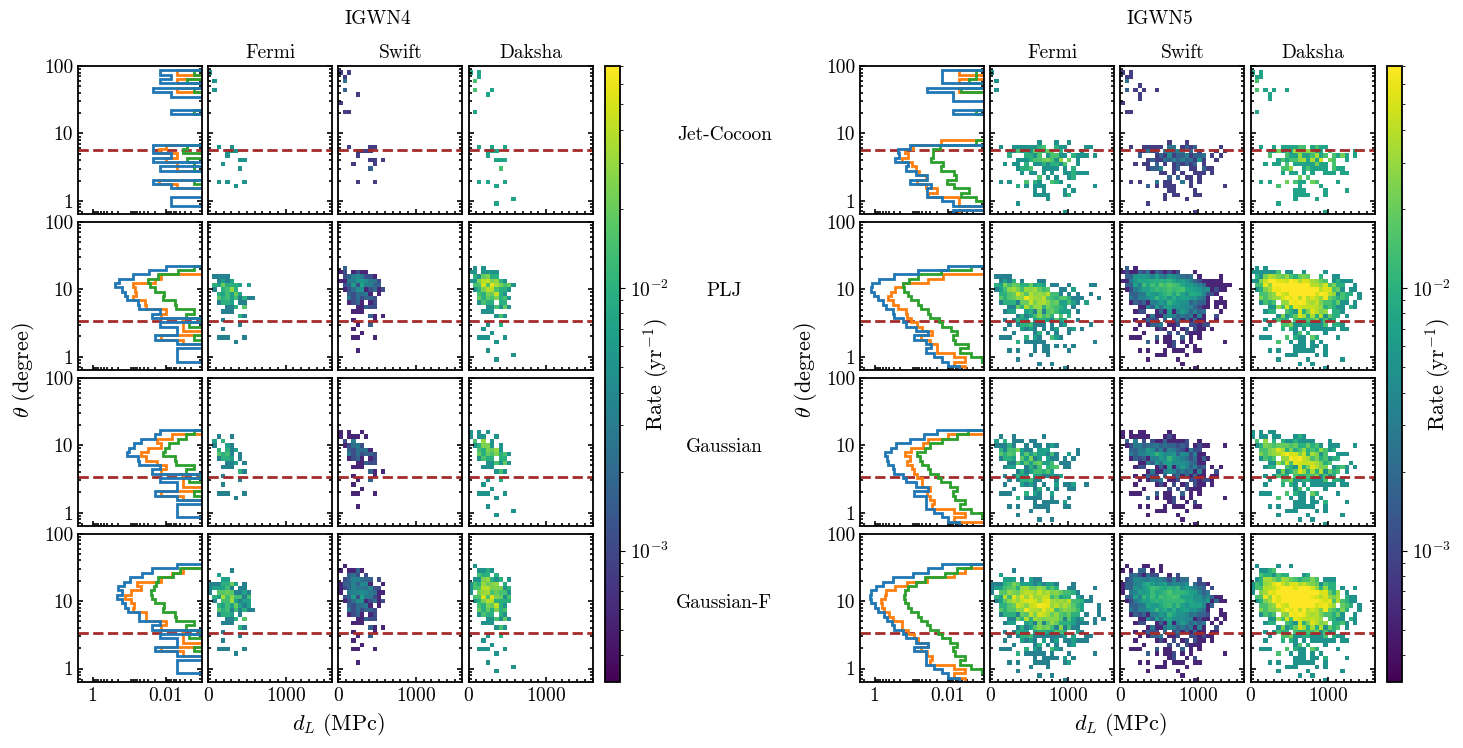}
    \caption{The distribution of the join sGRB and GW detected events in the viewing angle ($\theta_\mathrm{obs}$) and luminosity distance ($d_L$) space for IGWN4 (\textbf{Left}) and IGWN5 (\textbf{Right}) cases. Horizontal lines mark the jet opening angles ($\theta_0$). We span a larger region in $\theta-d_L$ space in IGWN5. With \emph{Daksha}, we span the largest space in $\theta-d_L$ with the highest efficiency and detection rates.}
    \label{fig:theta_dist_hist_rates_all_with_hist}
\end{figure*}

We now evaluate the rates of joint high-energy and gravitational wave detections of these events using the same criteria as defined above. As discussed before, we consider three missions: \emph{Fermi}, \emph{Swift}, and the upcoming twin-satellite \emph{Daksha}. For both detector networks that we consider (IGWN4 and IGWN5), we follow \citet{Petrov21} to define an ``above threshold'' detection criterion to be $\mathrm{SNR_{net}}=8$. On the other hand, if there is a confident EM detection of any event, the GW threshold can be lowered without compromising the false alarm rate, hence giving a significant joint detection. We refer to such events as ``sub-threshold'' events. While the computation of an exact network SNR cut-off value for sub-threshold events is beyond the scope of this work, we adopt a fiducial value of $\mathrm{SNR_{net,min}}=6.5$ for sub-threshold events. Considering the binary neutron star merger rate $R_{\mathrm{BNS}}=320\mathrm{\ GPc^{-3}~yr^{-1}}$ as discussed in \S\ref{subsec:comparing_satellites}, and the 30\% downtime for GW detectors (\S\ref{sec:sources_detectors}), our cutoffs lead to 15 (28) above-threshold events per year and 148 (264) sub-threshold events per year for IGWN4 (IGWN5), with the corresponding median luminosity distance of $\sim236 \rm{\ Mpc} \ (\sim 508 \rm{\ Mpc})$.

For the sensitivity of IGWN4, the overall joint sGRB + GW detection rates are rather low.
The total joint detection rates for the existing sGRB satellites \emph{Fermi} and \emph{Swift} lie in the range of $0.07-0.62\mathrm{~yr^{-1}}$ and $0.02-0.14\mathrm{~yr^{-1}}$ respectively, for the different models (Table~\ref{tab:rate_table}, Figure~\ref{fig:dist_cum_rate_both}). With an increased GW detection horizon with IGWN5, the detection rates increase to $0.8-4.0~\mathrm{yr^{-1}}$ and $0.15-1.0~\mathrm{yr^{-1}}$ for \emph{Fermi} and \emph{Swift}, respectively. For both IGWN4 and IGWN5, the fraction of GW events that are jointly detected with prompt emission, $f_{\mathrm{GW}}$, takes value in $0.5-4.0\%$ and $0.1-1.0\%$ for \emph{Fermi} and \emph{Swift} respectively. 

As expected, the rate of on-axis / off-axis detections is model-dependent. Off-axis events are geometrically more probable, and indeed most of the joint detections are expected to be off-axis ($\theta_\mathrm{obs} > \theta_0$) for the PLJ and Gaussian Jet models. For these models, fraction of jointly detected above-threshold off-axis events ($f_\mathrm{off}$) takes values in the range of $\sim60-95\%$ for \emph{Fermi} and $\sim80-95\%$ for \emph{Swift} respectively (Table~\ref{tab:rate_table}). However, the strong decline in off-axis luminosity for the Jet-Cocoon models makes these hard to detect, with $f_\mathrm{off}\sim50\%$ and $\sim15\%$ for IGWN4 and IGWN5, respectively. The sharp decline in $f_\mathrm{off}$ from IGWN4 to IGWN5 for Jet-Cocoon model originates from the domination of on-axis sGRB detections from the further away GW-detected events. The higher sensitivity of \emph{Swift} plays an important role here, as seen in Figure~\ref{fig:flux_histogram_all_models_satellites_scaled} --- a significant fraction of these off-axis events fall below the \emph{Fermi} detection threshold.

To consider a future perspective with a higher sensitivity mission, we examine the detection rates for \emph{Daksha}, which has a higher volumetric sensitivity for GW170817-like events than other missions \citep{Bhalerao22b}. As expected, the higher sensitivity and all-sky coverage yields much higher detection rates: $0.2-1.3\mathrm{\ yr^{-1}}$ for IGWN4, and $1.3-8.3\mathrm{\ yr^{-1}}$ for IGWN5: detecting $f_{\mathrm{GW}} = \sim1.0\%-8.4\%$ of the GW events. These rates are factors of $\sim2-9$ times higher than \emph{Fermi} or \emph{Swift}. The comparable sensitivities of \emph{Daksha} and \emph{Swift} lead to similar $f_{\mathrm{off}}$ values, but the increased sky coverage gives higher rates.

The region spanned by the jointly detected events in the $\theta_\mathrm{obs}-d_L$ space of the merger events increases significantly from IGWN4 to IGWN5 for all the satellites and models (Figure~\ref{fig:theta_dist_hist_rates_all_with_hist}). However, the overall increment in the joint detection rates, especially for the fainter off-axis and far away events, is limited by the satellite capability. For example, \emph{Swift} probes a larger region in $\theta-d_L$ space, i.e., can detect fainter ends of the events due to its better sensitivity. Nevertheless, the total number of events detected by \emph{Swift} is still not high due to its low sky coverage. \emph{Fermi}, on the other hand, probes much smaller region in this space owing to the higher flux threshold (i.e. lower sensitivity). However, with \emph{Daksha}, the detection rates will increase mainly within the viewing angle range of $5-20\degree$.  Such events will serve as a connecting link between cosmological on-axis sGRBs and GRB170817-like events, improving our understanding of the jet structure.
 
In addition to the satellites, the performance of GW detectors also limits joint detections. To quantify this, we consider the EM-detected sGRBs that are within the GW detection horizons, and calculate the fraction that is also detected in gravitational waves ($f_{\mathrm{sGRB}}$). We define the GW detection horizon to be the distance at which the cumulative rate attains $99\%$ of the total rate, which is $\approx523\mathrm{\ Mpc}$ and $\approx1127\mathrm{\ Mpc}$ for IGWN4 and IGWN5. For all the missions and emission models, $f_{\mathrm{sGRB}}$ takes values between $\sim40-50\%$ and $\sim55-65\%$ for IGWN4 and IGWN5 networks, respectively. These values are lower than $100\%$ due to both the non-zero downtime of the GW detectors and the limitation of detecting signals from the outer margins of the detection volume. The increment in the fraction from IGWN4 to IGWN5 results from improved GW detection efficiency.

\subsection{Comparing our rates with other works in the literature}\label{subsec:comparison}

In this section, we conduct a comparative analysis of our results with those from previous studies in the literature. We first compare the GW detection rates, followed by the sGRB+GW rates.

\subsubsection{GW detection rates}\label{subsubsec:gw}
With $R_{\mathrm{BNS}} = 320 \mathrm{~Gpc^{-3}~yr^{-1}}$ and detection threshold of $\mathrm{SNR}_{\rm det} = 8$, we obtain GW detection rates of $15 \mathrm{~yr^{-1}}$ ($123 \mathrm{~yr^{-1}}$) with IGWN4 (IGWN5) network. These rates align well with those constrained in other studies employing different population models within the uncertainty limits. For instance, \citet{Colombo22} obtain a rate of $7.7\mathrm{~yr^{-1}}$ with a similar $R_{\mathrm{BNS}}$, a stricter threshold detection criterion of $\mathrm{SNR_{net}}=12$, and $80\%$ detector duty cycle.

One notable difference is that our rates are less than half the value of $34~\mathrm{~yr^{-1}}$ obtained by \citet{Petrov21}. \citet{Kiendrebeogo23} obtain similar high rates: $36 \mathrm{~yr^{-1}}$ with $R_{\mathrm{BNS}} \approx 210 \mathrm{~GPc^{-3}~yr^{-1}}$ and $17 \mathrm{~yr^{-1}}$ with $R_{\mathrm{BNS}} \approx 170 \mathrm{~GPc^{-3}~yr^{-1}}$. While we have slightly different assumptions about the underlying astrophysical merger rate densities, the discrepancy can mainly be attributed to the methodology employed in estimating detectable sources. Both these works emulate the detection methods currently in use in LIGO/Virgo/KAGRA analyses. They employ similar population models as us but inject waveforms into simulated Gaussian noise, with detections based on matched filter SNR. In this scenario, noise can decrease or increase the signal strength, leading to false non-detections or chance detections respectively. Thanks to a uniform distribution of sources in volume, there are a larger number of events just below the detection threshold than above it: hence, this effect increases the number of events detected in their analyses. While these effects accurately mirror the actual data processing in IGWN, they overestimate the number of detections.

On the other hand, we base our estimations on the injected SNR of the sources, computed by simulating waveforms without noise and using power spectral density models. To test if this completely accounts for the discrepancy, we used the \citet{Kiendrebeogo23} data set to calculate the expected SNRs following our methods we found that the rates become consistent with our rates.

\subsubsection{Joint sGRB+GW detection rates}
Our sGRB+GW joint detection rates calculated in \S\ref{sec:emgw_detection} agree well with those estimated in several past works. Recent work by \citet{Colombo22}, provides a joint rate of $\sim0.17\mathrm{\ yr^{-1}}$ and $\sim0.03\mathrm{\ yr^{-1}}$ with \emph{Fermi} and \emph{Swift} respectively, which are consistent with with IGWN4 rates of this work. A lower fraction of successful jets ($f_{\mathrm{jet}}\sim52\%$) and stricter threshold detection criteria of $\mathrm{SNR_{net}}=12$ in their work leads to relatively lower rates. Our rates are also broadly consistent with \citet{Howell19} after appropriately scaling for the higher assumed $R_{\mathrm{BNS}}$ and $f_{\mathrm{jet}}=100\%$ used in their work. However, other works by \citet{Patricelli22} and \citet{Saleem20} yield significantly higher rates. This may arise from the brighter luminosity function or emission models. \citet{Patricelli22} included Gaussian noise of GW detectors, which, in light of the previous discussion about GW rates, can also be responsible for their relatively higher joint detection rate estimations. Our prediction of low values of $f_{\mathrm{GW}}$ is consistent with most past works \citep{Howell19,Colombo22,Patricelli22}. If we use GW detections following the \cite{Petrov21} or \cite{Kiendrebeogo23} method, we find that the GW+sGRB detection rates mentioned in Table~\ref{tab:rate_table} and elsewhere will approximately double.

\section{Discussion and Conclusion}\label{sec:conclusion}

\begin{figure}
    \centering
    \includegraphics[width=\linewidth]{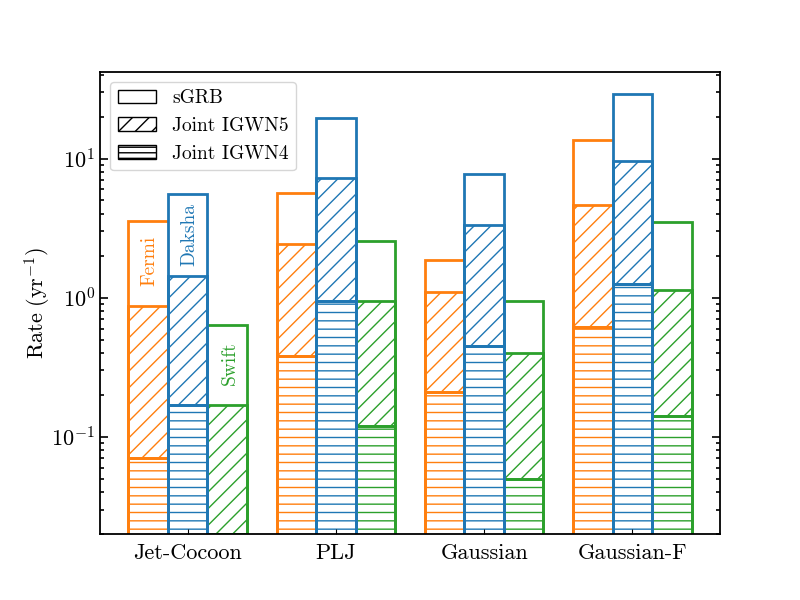}
    \caption{The sGRB and joint sGRB and GW rates for the three missions with different prompt emission models. The rates are computed for $R_{\mathrm{BNS}}=320\mathrm{\ GPc^{-3}~yr^{-1}}$. The sGRB rate is only for events with $d_L<1.6\mathrm{\ Gpc}$, the distance limit of this work.}
    \label{fig:summary_plot}
\end{figure}

In this work, we examine the prospects of detecting sGRB counterparts and joint sGRB and GW detection of BNS merger events by three high-energy satellites: the existing missions \emph{Fermi} and \emph{Swift} and the upcoming all-sky high sensitivity mission \emph{Daksha}; for two GW detector networks: current sensitivity 4-detector (HLVK, IGWN4 case) and future A+ sensitivity 5-detector (HLVKI, IGWN5 case) network. Many of the jointly detected events are observed beyond the jet core (off-axis). Hence, we use several jet models predicting off-axis emission to calculate the joint detection rates (power law jet and jet-cocoon models by \cite{Beniamini19a} and Gaussian jet model by \citet{Ioka19}, see \S\ref{sec:models}). In addition, to test the effect of the assumed luminosity function on the predicted rates, we use two different luminosity functions, given by \citet{WP15} and \citet{Fong15}, for one of the representative models (the Gaussian model, see \S\ref{sec:sims}).

We use the volumetric BNS merger rate  $R_{\mathrm{BNS}}=320\mathrm{~GPc^{-3}~yr^{-1}}$ \citep{Abbott21b, Abbott21c} for calculating joint detector rates. Our predicted joint detection rates for the existing satellites \emph{Fermi} and \emph{Swift} with IGWN4 (IGWN5) are 0.07-0.62$\mathrm{\ yr^{-1}}$ (0.8-4.0$\mathrm{\ yr^{-1}}$) and 0.02-0.14$\mathrm{\ yr^{-1}}$ (0.15-1.0$\mathrm{\ yr^{-1}}$), respectively. These rates increase by a factor of $\sim2-9$ for the proposed \emph{Daksha} mission. The predicted joint detection rates with \emph{Daksha} with IGWN4 (IGWN5) network and GW detection over threshold SNR lie in the range of $0.2-1.3\mathrm{\ yr^{-1}}$ ($1.3-8.3\mathrm{\ yr^{-1}}$). This highlights the need for a more sensitive future mission:
The IGWN5 rates imply that \emph{Daksha} will lead to at least one joint sGRB and GW detection from BNS merger event per year, which is a notable improvement from no event in $6$ years (since GRB170817A). Figure \ref{fig:summary_plot} summarizes the results of this work (also see Table~\ref{tab:rate_table}). We note here that our rate estimations with optimal SNRs of the injected BNS merger events are conservative. Comparison with the works of \citet{Petrov21} and \citet{Kiendrebeogo23} suggests that Gaussian noise in GW detectors can assist the detection of otherwise GW-faint events, which can potentially increase both the GW and sGRB+GW detection rates by a factor of two (see \S\ref{subsec:comparison}).

We show that with future GW detection networks (like IGWN5), joint detection rates at lower viewing angles ($\theta_\mathrm{obs}\lessapprox2\theta_0$, $\theta_0$ being the jet opening angle) increase significantly from IGWN4 to IGWN5. Larger angle detections are, however, limited by satellite performance. Nevertheless, within the detectable viewing angle range, \emph{Daksha} is expected to perform better than the existing satellites. It will probe the largest region in the space of viewing angle and distance of the merger events with the highest efficiency among the satellites (see Figure \ref{fig:theta_dist_hist_rates_all_with_hist}). The predicted range of off-axis joint detection rate with \emph{Daksha} and IGWN4 (IGWN5) of $0.09-1.2\mathrm{\ yr^{-1}}$ ($0.2-7.77\mathrm{\ yr^{-1}}$), is higher than that of both \emph{Swift} and \emph{Fermi} by factors of $2-6$. This underscores the need for future missions that combine high sensitivity with all-sky coverage for detecting and characterizing these events, and in turn understanding the physics of compact object mergers and the post-merger radiative processes.

\section*{Acknowledgements}

We thank the anonymous referee for the constructive comments. We thank Michael Coughlin, Leo Singer, and Weizmann Kiendrebeogo profusely for their help, time and effort in understanding the differences between their rates and ours. AP's research is supported by SERB-Power fellowship grant SPF/2021/000036, DST, India. The research of PB is supported by the United States-Israel Binational Science Foundation (BSF) under grant no. 2020747

\section*{Data Availability}

The injection data and all the derived quantities and models produced in this study will be shared on reasonable request to the corresponding author.



\bibliographystyle{mnras}
\bibliography{example} 

\begin{thebibliography}{}
\makeatletter
\relax
\def\mn@urlcharsother{\let\do\@makeother \do\$\do\&\do\#\do\^\do\_\do\%\do\~}
\def\mn@doi{\begingroup\mn@urlcharsother \@ifnextchar [ {\mn@doi@}
  {\mn@doi@[]}}
\def\mn@doi@[#1]#2{\def\@tempa{#1}\ifx\@tempa\@empty \href
  {http://dx.doi.org/#2} {doi:#2}\else \href {http://dx.doi.org/#2} {#1}\fi
  \endgroup}
\def\mn@eprint#1#2{\mn@eprint@#1:#2::\@nil}
\def\mn@eprint@arXiv#1{\href {http://arxiv.org/abs/#1} {{\tt arXiv:#1}}}
\def\mn@eprint@dblp#1{\href {http://dblp.uni-trier.de/rec/bibtex/#1.xml}
  {dblp:#1}}
\def\mn@eprint@#1:#2:#3:#4\@nil{\def\@tempa {#1}\def\@tempb {#2}\def\@tempc
  {#3}\ifx \@tempc \@empty \let \@tempc \@tempb \let \@tempb \@tempa \fi \ifx
  \@tempb \@empty \def\@tempb {arXiv}\fi \@ifundefined
  {mn@eprint@\@tempb}{\@tempb:\@tempc}{\expandafter \expandafter \csname
  mn@eprint@\@tempb\endcsname \expandafter{\@tempc}}}

\bibitem[\protect\citeauthoryear{{Abbott} et~al.,}{{Abbott}
  et~al.}{2017}]{Abbott17}
{Abbott} B.~P.,  et~al., 2017, \mn@doi [\apjl] {10.3847/2041-8213/aa920c},
  \href {https://ui.adsabs.harvard.edu/abs/2017ApJ...848L..13A} {848, L13}

\bibitem[\protect\citeauthoryear{{Abbott} et~al.,}{{Abbott}
  et~al.}{2021a}]{Abbott21c}
{Abbott} R.,  et~al., 2021a, \mn@doi [\apjl] {10.3847/2041-8213/abe949}, \href
  {https://ui.adsabs.harvard.edu/abs/2021ApJ...913L...7A} {913, L7}

\bibitem[\protect\citeauthoryear{{Abbott} et~al.,}{{Abbott}
  et~al.}{2021b}]{Abbott21}
{Abbott} R.,  et~al., 2021b, \mn@doi [\apjl] {10.3847/2041-8213/abffcd}, \href
  {https://ui.adsabs.harvard.edu/abs/2021ApJ...913L..27A} {913, L27}

\bibitem[\protect\citeauthoryear{{Abbott} et~al.,}{{Abbott}
  et~al.}{2021c}]{Abbott21b}
{Abbott} R.,  et~al., 2021c, \mn@doi [\apjl] {10.3847/2041-8213/ac082e}, \href
  {https://ui.adsabs.harvard.edu/abs/2021ApJ...915L...5A} {915, L5}

\bibitem[\protect\citeauthoryear{{Acernese} et~al.,}{{Acernese}
  et~al.}{2015}]{Acernese15}
{Acernese} F.,  et~al., 2015, \mn@doi [Classical and Quantum Gravity]
  {10.1088/0264-9381/32/2/024001}, \href
  {https://ui.adsabs.harvard.edu/abs/2015CQGra..32b4001A} {32, 024001}

\bibitem[\protect\citeauthoryear{{Alexander} et~al.,}{{Alexander}
  et~al.}{2018}]{Alexander18}
{Alexander} K.~D.,  et~al., 2018, \mn@doi [\apjl] {10.3847/2041-8213/aad637},
  \href {https://ui.adsabs.harvard.edu/abs/2018ApJ...863L..18A} {863, L18}

\bibitem[\protect\citeauthoryear{{Aso}, {Michimura}, {Somiya}, {Ando},
  {Miyakawa}, {Sekiguchi}, {Tatsumi}  \& {Yamamoto}}{{Aso}
  et~al.}{2013}]{Aso13}
{Aso} Y.,  {Michimura} Y.,  {Somiya} K.,  {Ando} M.,  {Miyakawa} O.,
  {Sekiguchi} T.,  {Tatsumi} D.,   {Yamamoto} H.,  2013, \mn@doi [\prd]
  {10.1103/PhysRevD.88.043007}, \href
  {https://ui.adsabs.harvard.edu/abs/2013PhRvD..88d3007A} {88, 043007}

\bibitem[\protect\citeauthoryear{{Band} et~al.,}{{Band} et~al.}{1993}]{Band93}
{Band} D.,  et~al., 1993, \mn@doi [\apj] {10.1086/172995}, \href
  {https://ui.adsabs.harvard.edu/abs/1993ApJ...413..281B} {413, 281}

\bibitem[\protect\citeauthoryear{{Barsotti}, {McCuller}, {Evans}  \&
  {Fritschel}}{{Barsotti} et~al.}{2018}]{Barsotti18}
{Barsotti} L.,  {McCuller} L.,  {Evans} M.,   {Fritschel} P.,  2018, The A+
  design curve, \url {https://dcc.ligo.org/LIGO-T1800042-v4/public}

\bibitem[\protect\citeauthoryear{{Barthelmy} et~al.,}{{Barthelmy}
  et~al.}{2005}]{Barthelmy05}
{Barthelmy} S.~D.,  et~al., 2005, \mn@doi [\ssr] {10.1007/s11214-005-5096-3},
  \href {https://ui.adsabs.harvard.edu/abs/2005SSRv..120..143B} {120, 143}

\bibitem[\protect\citeauthoryear{{Beniamini} \& {Nakar}}{{Beniamini} \&
  {Nakar}}{2019}]{Beniamini19b}
{Beniamini} P.,  {Nakar} E.,  2019, \mn@doi [\mnras] {10.1093/mnras/sty3110},
  \href {https://ui.adsabs.harvard.edu/abs/2019MNRAS.482.5430B} {482, 5430}

\bibitem[\protect\citeauthoryear{{Beniamini}, {Petropoulou}, {Barniol Duran}
  \& {Giannios}}{{Beniamini} et~al.}{2019}]{Beniamini19a}
{Beniamini} P.,  {Petropoulou} M.,  {Barniol Duran} R.,   {Giannios} D.,  2019,
  \mn@doi [\mnras] {10.1093/mnras/sty3093}, \href
  {https://ui.adsabs.harvard.edu/abs/2019MNRAS.483..840B} {483, 840}

\bibitem[\protect\citeauthoryear{{Beniamini}, {Granot}  \& {Gill}}{{Beniamini}
  et~al.}{2020a}]{Beniamini2020}
{Beniamini} P.,  {Granot} J.,   {Gill} R.,  2020a, \mn@doi [\mnras]
  {10.1093/mnras/staa538}, \href
  {https://ui.adsabs.harvard.edu/abs/2020MNRAS.493.3521B} {493, 3521}

\bibitem[\protect\citeauthoryear{{Beniamini}, {Duran}, {Petropoulou}  \&
  {Giannios}}{{Beniamini} et~al.}{2020b}]{BDPG2020}
{Beniamini} P.,  {Duran} R.~B.,  {Petropoulou} M.,   {Giannios} D.,  2020b,
  \mn@doi [\apjl] {10.3847/2041-8213/ab9223}, \href
  {https://ui.adsabs.harvard.edu/abs/2020ApJ...895L..33B} {895, L33}

\bibitem[\protect\citeauthoryear{{Beniamini}, {Gill}  \& {Granot}}{{Beniamini}
  et~al.}{2022}]{Beniamini2022}
{Beniamini} P.,  {Gill} R.,   {Granot} J.,  2022, \mn@doi [\mnras]
  {10.1093/mnras/stac1821}, \href
  {https://ui.adsabs.harvard.edu/abs/2022MNRAS.515..555B} {515, 555}

\bibitem[\protect\citeauthoryear{{Berger}}{{Berger}}{2014}]{Berger14}
{Berger} E.,  2014, \mn@doi [\araa] {10.1146/annurev-astro-081913-035926},
  \href {https://ui.adsabs.harvard.edu/abs/2014ARA&A..52...43B} {52, 43}

\bibitem[\protect\citeauthoryear{{Bhalerao} et~al.,}{{Bhalerao}
  et~al.}{2022a}]{Bhalerao22a}
{Bhalerao} V.,  et~al., 2022a, \mn@doi [arXiv e-prints]
  {10.48550/arXiv.2211.12052}, \href
  {https://ui.adsabs.harvard.edu/abs/2022arXiv221112052B} {p. arXiv:2211.12052}

\bibitem[\protect\citeauthoryear{{Bhalerao} et~al.,}{{Bhalerao}
  et~al.}{2022b}]{Bhalerao22b}
{Bhalerao} V.,  et~al., 2022b, \mn@doi [arXiv e-prints]
  {10.48550/arXiv.2211.12055}, \href
  {https://ui.adsabs.harvard.edu/abs/2022arXiv221112055B} {p. arXiv:2211.12055}

\bibitem[\protect\citeauthoryear{{Chen}, {Holz}, {Miller}, {Evans}, {Vitale}
  \& {Creighton}}{{Chen} et~al.}{2021}]{Chen21}
{Chen} H.-Y.,  {Holz} D.~E.,  {Miller} J.,  {Evans} M.,  {Vitale} S.,
  {Creighton} J.,  2021, \mn@doi [Classical and Quantum Gravity]
  {10.1088/1361-6382/abd594}, \href
  {https://ui.adsabs.harvard.edu/abs/2021CQGra..38e5010C} {38, 055010}

\bibitem[\protect\citeauthoryear{{Colombo}, {Salafia}, {Gabrielli},
  {Ghirlanda}, {Giacomazzo}, {Perego}  \& {Colpi}}{{Colombo}
  et~al.}{2022}]{Colombo22}
{Colombo} A.,  {Salafia} O.~S.,  {Gabrielli} F.,  {Ghirlanda} G.,  {Giacomazzo}
  B.,  {Perego} A.,   {Colpi} M.,  2022, \mn@doi [\apj]
  {10.3847/1538-4357/ac8d00}, \href
  {https://ui.adsabs.harvard.edu/abs/2022ApJ...937...79C} {937, 79}

\bibitem[\protect\citeauthoryear{{Fong}, {Berger}, {Margutti}  \&
  {Zauderer}}{{Fong} et~al.}{2015}]{Fong15}
{Fong} W.,  {Berger} E.,  {Margutti} R.,   {Zauderer} B.~A.,  2015, \mn@doi
  [\apj] {10.1088/0004-637X/815/2/102}, \href
  {https://ui.adsabs.harvard.edu/abs/2015ApJ...815..102F} {815, 102}

\bibitem[\protect\citeauthoryear{{Ghirlanda} et~al.,}{{Ghirlanda}
  et~al.}{2019}]{Ghirlanda19}
{Ghirlanda} G.,  et~al., 2019, \mn@doi [Science] {10.1126/science.aau8815},
  \href {https://ui.adsabs.harvard.edu/abs/2019Sci...363..968G} {363, 968}

\bibitem[\protect\citeauthoryear{{Goldstein} et~al.,}{{Goldstein}
  et~al.}{2017}]{Goldstein17}
{Goldstein} A.,  et~al., 2017, \mn@doi [\apjl] {10.3847/2041-8213/aa8319},
  \href {https://ui.adsabs.harvard.edu/abs/2017ApJ...846L...5G} {846, L5}

\bibitem[\protect\citeauthoryear{Hannam, Schmidt, Boh\'e, Haegel, Husa, Ohme,
  Pratten  \& P\"urrer}{Hannam et~al.}{2014}]{Hannam14}
Hannam M.,  Schmidt P.,  Boh\'e A.,  Haegel L.,  Husa S.,  Ohme F.,  Pratten
  G.,   P\"urrer M.,  2014, \mn@doi [Phys. Rev. Lett.]
  {10.1103/PhysRevLett.113.151101}, 113, 151101

\bibitem[\protect\citeauthoryear{{Howell}, {Ackley}, {Rowlinson}  \&
  {Coward}}{{Howell} et~al.}{2019}]{Howell19}
{Howell} E.~J.,  {Ackley} K.,  {Rowlinson} A.,   {Coward} D.,  2019, \mn@doi
  [\mnras] {10.1093/mnras/stz455}, \href
  {https://ui.adsabs.harvard.edu/abs/2019MNRAS.485.1435H} {485, 1435}

\bibitem[\protect\citeauthoryear{{Ioka} \& {Nakamura}}{{Ioka} \&
  {Nakamura}}{2019}]{Ioka19}
{Ioka} K.,  {Nakamura} T.,  2019, \mn@doi [\mnras] {10.1093/mnras/stz1650},
  \href {https://ui.adsabs.harvard.edu/abs/2019MNRAS.487.4884I} {487, 4884}

\bibitem[\protect\citeauthoryear{{Iyer}, {Souradeep}, {Unnikrishnan},
  {Dhurandhar}, {Raja}  \& {Sengupta}}{{Iyer} et~al.}{2011}]{Iyer11}
{Iyer} B.,  {Souradeep} T.,  {Unnikrishnan} C.,  {Dhurandhar} S.,  {Raja} S.,
  {Sengupta} A.,  2011, LIGO-India, Proposal of the Consortium for Indian
  Initiative in Gravitational-wave Observations (IndIGO), \url
  {https://dcc.ligo.org/LIGO-M1100296/public}

\bibitem[\protect\citeauthoryear{{Kaneko}, {Preece}, {Briggs}, {Paciesas},
  {Meegan}  \& {Band}}{{Kaneko} et~al.}{2006}]{Kaneko06}
{Kaneko} Y.,  {Preece} R.~D.,  {Briggs} M.~S.,  {Paciesas} W.~S.,  {Meegan}
  C.~A.,   {Band} D.~L.,  2006, \mn@doi [\apjs] {10.1086/505911}, \href
  {https://ui.adsabs.harvard.edu/abs/2006ApJS..166..298K} {166, 298}

\bibitem[\protect\citeauthoryear{{Khan}, {Husa}, {Hannam}, {Ohme},
  {P{\"u}rrer}, {Forteza}  \& {Boh{\'e}}}{{Khan} et~al.}{2016}]{Khan16}
{Khan} S.,  {Husa} S.,  {Hannam} M.,  {Ohme} F.,  {P{\"u}rrer} M.,  {Forteza}
  X.~J.,   {Boh{\'e}} A.,  2016, \mn@doi [\prd] {10.1103/PhysRevD.93.044007},
  \href {https://ui.adsabs.harvard.edu/abs/2016PhRvD..93d4007K} {93, 044007}

\bibitem[\protect\citeauthoryear{{Kiendrebeogo} et~al.,}{{Kiendrebeogo}
  et~al.}{2023}]{Kiendrebeogo23}
{Kiendrebeogo} R.~W.,  et~al., 2023, \mn@doi [\apj] {10.3847/1538-4357/acfcb1},
  \href {https://ui.adsabs.harvard.edu/abs/2023ApJ...958..158K} {958, 158}

\bibitem[\protect\citeauthoryear{{Kouveliotou}, {Meegan}, {Fishman}, {Bhat},
  {Briggs}, {Koshut}, {Paciesas}  \& {Pendleton}}{{Kouveliotou}
  et~al.}{1993}]{Kouveliotou93}
{Kouveliotou} C.,  {Meegan} C.~A.,  {Fishman} G.~J.,  {Bhat} N.~P.,  {Briggs}
  M.~S.,  {Koshut} T.~M.,  {Paciesas} W.~S.,   {Pendleton} G.~N.,  1993,
  \mn@doi [\apjl] {10.1086/186969}, \href
  {https://ui.adsabs.harvard.edu/abs/1993ApJ...413L.101K} {413, L101}

\bibitem[\protect\citeauthoryear{{LIGO Scientific Collaboration} et~al.,}{{LIGO
  Scientific Collaboration} et~al.}{2015}]{Aasi15}
{LIGO Scientific Collaboration} et~al., 2015, \mn@doi [Classical and Quantum
  Gravity] {10.1088/0264-9381/32/7/074001}, \href
  {https://ui.adsabs.harvard.edu/abs/2015CQGra..32g4001L} {32, 074001}

\bibitem[\protect\citeauthoryear{{Lien} et~al.,}{{Lien} et~al.}{2014}]{Lien14}
{Lien} A.,  et~al., 2014, in Proceedings of Swift: 10 Years of Discovery (SWIFT
  10. p.~38, \mn@doi{10.22323/1.233.0038}

\bibitem[\protect\citeauthoryear{{Lyman} et~al.,}{{Lyman}
  et~al.}{2018}]{Lyman18}
{Lyman} J.~D.,  et~al., 2018, \mn@doi [Nature Astronomy]
  {10.1038/s41550-018-0511-3}, \href
  {https://ui.adsabs.harvard.edu/abs/2018NatAs...2..751L} {2, 751}

\bibitem[\protect\citeauthoryear{{Margutti} et~al.,}{{Margutti}
  et~al.}{2018}]{Margutti18}
{Margutti} R.,  et~al., 2018, \mn@doi [\apjl] {10.3847/2041-8213/aab2ad}, \href
  {https://ui.adsabs.harvard.edu/abs/2018ApJ...856L..18M} {856, L18}

\bibitem[\protect\citeauthoryear{{Mohan}, {Saleem}  \& {Resmi}}{{Mohan}
  et~al.}{2022}]{Sreelekshmi}
{Mohan} S.,  {Saleem} M.,   {Resmi} L.,  2022, \mn@doi [\mnras]
  {10.1093/mnras/stab3120}, \href
  {https://ui.adsabs.harvard.edu/abs/2022MNRAS.511.2356M} {511, 2356}

\bibitem[\protect\citeauthoryear{{Mooley} et~al.,}{{Mooley}
  et~al.}{2018a}]{Mooley18a}
{Mooley} K.~P.,  et~al., 2018a, \mn@doi [\nat] {10.1038/s41586-018-0486-3},
  \href {https://ui.adsabs.harvard.edu/abs/2018Natur.561..355M} {561, 355}

\bibitem[\protect\citeauthoryear{{Mooley} et~al.,}{{Mooley}
  et~al.}{2018b}]{Mooley18b}
{Mooley} K.~P.,  et~al., 2018b, \mn@doi [\apjl] {10.3847/2041-8213/aaeda7},
  \href {https://ui.adsabs.harvard.edu/abs/2018ApJ...868L..11M} {868, L11}

\bibitem[\protect\citeauthoryear{{Nakar}}{{Nakar}}{2007}]{Nakar07}
{Nakar} E.,  2007, \mn@doi [\physrep] {10.1016/j.physrep.2007.02.005}, \href
  {https://ui.adsabs.harvard.edu/abs/2007PhR...442..166N} {442, 166}

\bibitem[\protect\citeauthoryear{{Nakar}}{{Nakar}}{2020}]{Nakar20}
{Nakar} E.,  2020, \mn@doi [\physrep] {10.1016/j.physrep.2020.08.008}, \href
  {https://ui.adsabs.harvard.edu/abs/2020PhR...886....1N} {886, 1}

\bibitem[\protect\citeauthoryear{{Nava}, {Ghirlanda}, {Ghisellini}  \&
  {Celotti}}{{Nava} et~al.}{2011}]{Nava2011}
{Nava} L.,  {Ghirlanda} G.,  {Ghisellini} G.,   {Celotti} A.,  2011, \mn@doi
  [\aap] {10.1051/0004-6361/201016270}, \href
  {https://ui.adsabs.harvard.edu/abs/2011A&A...530A..21N} {530, A21}

\bibitem[\protect\citeauthoryear{{{\"O}zel} \& {Freire}}{{{\"O}zel} \&
  {Freire}}{2016}]{ozel16}
{{\"O}zel} F.,  {Freire} P.,  2016, \mn@doi [\araa]
  {10.1146/annurev-astro-081915-023322}, \href
  {https://ui.adsabs.harvard.edu/abs/2016ARA&A..54..401O} {54, 401}

\bibitem[\protect\citeauthoryear{Pai, Dhurandhar  \& Bose}{Pai
  et~al.}{2001}]{Pai:2000zt}
Pai A.,  Dhurandhar S.,   Bose S.,  2001, \mn@doi [Phys. Rev.]
  {10.1103/PhysRevD.64.042004}, D64, 042004

\bibitem[\protect\citeauthoryear{{Patricelli}, {Bernardini}, {Mapelli},
  {D'Avanzo}, {Santoliquido}, {Cella}, {Razzano}  \& {Cuoco}}{{Patricelli}
  et~al.}{2022}]{Patricelli22}
{Patricelli} B.,  {Bernardini} M.~G.,  {Mapelli} M.,  {D'Avanzo} P.,
  {Santoliquido} F.,  {Cella} G.,  {Razzano} M.,   {Cuoco} E.,  2022, \mn@doi
  [\mnras] {10.1093/mnras/stac1167}, \href
  {https://ui.adsabs.harvard.edu/abs/2022MNRAS.513.4159P} {513, 4159}

\bibitem[\protect\citeauthoryear{{Petrov} et~al.,}{{Petrov}
  et~al.}{2022}]{Petrov21}
{Petrov} P.,  et~al., 2022, \mn@doi [\apj] {10.3847/1538-4357/ac366d}, \href
  {https://ui.adsabs.harvard.edu/abs/2022ApJ...924...54P} {924, 54}

\bibitem[\protect\citeauthoryear{{Planck Collaboration} et~al.,}{{Planck
  Collaboration} et~al.}{2020}]{Planck13}
{Planck Collaboration} et~al., 2020, \mn@doi [\aap]
  {10.1051/0004-6361/201833910}, \href
  {https://ui.adsabs.harvard.edu/abs/2020A&A...641A...6P} {641, A6}

\bibitem[\protect\citeauthoryear{{Poolakkil} et~al.,}{{Poolakkil}
  et~al.}{2021}]{Poolakkil21}
{Poolakkil} S.,  et~al., 2021, \mn@doi [\apj] {10.3847/1538-4357/abf24d}, \href
  {https://ui.adsabs.harvard.edu/abs/2021ApJ...913...60P} {913, 60}

\bibitem[\protect\citeauthoryear{{Saleem}}{{Saleem}}{2020}]{Saleem20}
{Saleem} M.,  2020, \mn@doi [\mnras] {10.1093/mnras/staa303}, \href
  {https://ui.adsabs.harvard.edu/abs/2020MNRAS.493.1633S} {493, 1633}

\bibitem[\protect\citeauthoryear{{Saleem} et~al.,}{{Saleem}
  et~al.}{2022}]{Saleem22}
{Saleem} M.,  et~al., 2022, \mn@doi [Classical and Quantum Gravity]
  {10.1088/1361-6382/ac3b99}, \href
  {https://ui.adsabs.harvard.edu/abs/2022CQGra..39b5004S} {39, 025004}

\bibitem[\protect\citeauthoryear{{Savchenko} et~al.,}{{Savchenko}
  et~al.}{2017}]{Savchenko17}
{Savchenko} V.,  et~al., 2017, \mn@doi [\apjl] {10.3847/2041-8213/aa8f94},
  \href {https://ui.adsabs.harvard.edu/abs/2017ApJ...848L..15S} {848, L15}

\bibitem[\protect\citeauthoryear{{Schmidt}, {Hannam}  \& {Husa}}{{Schmidt}
  et~al.}{2012}]{Schmidt12}
{Schmidt} P.,  {Hannam} M.,   {Husa} S.,  2012, \mn@doi [\prd]
  {10.1103/PhysRevD.86.104063}, \href
  {https://ui.adsabs.harvard.edu/abs/2012PhRvD..86j4063S} {86, 104063}

\bibitem[\protect\citeauthoryear{{Troja} et~al.,}{{Troja}
  et~al.}{2017}]{Troja17}
{Troja} E.,  et~al., 2017, \mn@doi [\nat] {10.1038/nature24290}, \href
  {https://ui.adsabs.harvard.edu/abs/2017Natur.551...71T} {551, 71}

\bibitem[\protect\citeauthoryear{{Troja} et~al.,}{{Troja}
  et~al.}{2018}]{Troja18}
{Troja} E.,  et~al., 2018, \mn@doi [Nature Communications]
  {10.1038/s41467-018-06558-7}, \href
  {https://ui.adsabs.harvard.edu/abs/2018NatCo...9.4089T} {9, 4089}

\bibitem[\protect\citeauthoryear{{Troja} et~al.,}{{Troja}
  et~al.}{2020}]{Troja20}
{Troja} E.,  et~al., 2020, \mn@doi [\mnras] {10.1093/mnras/staa2626}, \href
  {https://ui.adsabs.harvard.edu/abs/2020MNRAS.498.5643T} {498, 5643}

\bibitem[\protect\citeauthoryear{{Wanderman} \& {Piran}}{{Wanderman} \&
  {Piran}}{2015}]{WP15}
{Wanderman} D.,  {Piran} T.,  2015, \mn@doi [\mnras] {10.1093/mnras/stv123},
  \href {https://ui.adsabs.harvard.edu/abs/2015MNRAS.448.3026W} {448, 3026}

\bibitem[\protect\citeauthoryear{{Zhu}, {Thrane}, {Os{\l}owski}, {Levin}  \&
  {Lasky}}{{Zhu} et~al.}{2018}]{Zhu18}
{Zhu} X.,  {Thrane} E.,  {Os{\l}owski} S.,  {Levin} Y.,   {Lasky} P.~D.,  2018,
  \mn@doi [\prd] {10.1103/PhysRevD.98.043002}, \href
  {https://ui.adsabs.harvard.edu/abs/2018PhRvD..98d3002Z} {98, 043002}

\bibitem[\protect\citeauthoryear{{von Kienlin} et~al.,}{{von Kienlin}
  et~al.}{2020}]{vKienlin20}
{von Kienlin} A.,  et~al., 2020, \mn@doi [\apj] {10.3847/1538-4357/ab7a18},
  \href {https://ui.adsabs.harvard.edu/abs/2020ApJ...893...46V} {893, 46}

\makeatother
\end{thebibliography}





\bsp	
\label{lastpage}
\end{document}